# Semi-Automatic Indexing of Multilingual Documents

Ulrich Schiel[1], Ianna M. Sodré Ferreira de Sousa[1], Edberto Ferneda[2]

[1]Universidade Federal da Paraíba, Departamento de Sistemas e Computação
Caixa Postal 10106, 58.109-970 - Campina Grande – PB / BRAZIL
[2]Escola Técnica Federal de Alagoas, Processamento de Dados
Rua Barão de Atalaia s/n, 57.020-510 - Maceió –AL / BRAZIL

[1](ulrich,ianna)@dsc.ufpb.br, [2]edberto@etfal.g12.br,

## ABSTRACT

With the growing significance of digital libraries and the Internet, more and more electronic texts become accessible to a wide and geographically disperse public. This requires adequate tools to facilitate indexing, storage, and retrieval of documents written in different languages. We present a method for semi-automatic indexing of electronic documents and construction of a multilingual thesaurus, which can be used for query formulation and information retrieval. We use special dictionaries and user interaction in order to solve ambiguities and find adequate canonical terms in the language and adequate abstract language-independent terms. The abstract thesaurus is updated incrementally by new indexed documents and is used to search document concerning terms in a query to the document base.

## 1. Introduction

In order to get extensive and precise retrieval of textual information, correct and consistent analysis of incoming documents is necessary. The most broadly-used technique for this analysis is indexing. An index file becomes an intermediate representation between a query and the document base.

One of the most popular structures for complex indexes is a semantic net called thesaurus. The nodes are single or composite terms and the links are semantic relationships between these terms.

Bruandet [4,5] has developed an automatic indexing technique which was extended by Gammoudi [6] for optimal thesaurus generation for a given set of documents. The nodes are rectangles where the left side contains a set of terms and the right side the set of documents indexing the terms. Higher rectangles in the thesaurus contains more specific term sets and less documents. One extension of this technique is the algorithm of Pinto [14] which permits an incremental addition of index terms of new incoming documents, updating the Thesaurus.

We show in this paper how this extended version of Gammoudi's technique can be used in an environment with multilingual documents and queries whose language need not be the same as that of the searched documents. The main ideas are the using of monolingual dictionaries in order to, with the user's help, eliminate ambiguities, and an abstract, language independent thesaurus. The terms of a query are also converted to the abstract terms in order to find the corresponding documents.

Whereas most work on Multilingual (or Translingual) Information Retrieval concentrate on query processing [16], our focus is on the indexing of digital documents.

In the next section we give basic mathematical definitions and notions of thesauri, including our definition of a multilingual rectangular thesaurus. The following section 3 shows the procedure of term extraction from documents, finding the abstract concept and the term-document association and inclusion of the new rectangle in the existing rectangular thesaurus. Section 4 shows the query and retrieval environment and, finally, section 5 discusses some related work and concludes the paper.

## 2. Rectangular Thesaurus: Basic Concepts

The main concept used for thesaurus construction is the *binary relation*. A binary relation can be decomposed in a minimal set of optimal rectangles by the method of *Rectangular Decomposition of a Binary Relation* **[6, 3]**. The extraction of rectangles from a finite binary relation has been extensively studied in the context of Lattice Theory and has proven to be very useful in many computer science applications..

A rectangle of a binary relation R is a pair of sets (A, B) such that $A \times B \subseteq R$. More precisely

### Definition 1: Rectangle

Let *R* be a binary relation defined from *E* to *F*. A rectangle of *R* is a pair of sets *(A,B)* such that $A \subseteq E$, $B \subseteq F$ and $A \times B \subseteq R$. *A* is the domain of the rectangle where *B* is the codomain.

A rectangle *(A,B)* of a relation *R* is **maximal** iff, for each rectangle *(A',B')*

$A \times B \subseteq A' \times B' \subseteq R \rightarrow A = A'$ e $B = B'$.

A binary relation can be represented by different sets of rectangles. In order to gain storage space the following coefficient is important for the selection of an optimal set of rectangles representing the relation.

### Definition 2: Gain in storage space

The **gain** in storage space of a rectangle $RE=(A,B)$ is given by:

$g(RE) = [Card(A) \times Card(B)] - [Card(A) + Card(B)]$

where Card(A) is the cardinality of the set A.

The gain becomes significant if Card(A) > 2 and Card(B) > 2, then g(RE) > 0 and grows with Card(A) and Card(B). On the other hand, there is no gain (g(GE)<0) if Card(A) =1 or Card(B) =1. The notion of gain is very important because it allows us to save memory space, in particular when we need to store a large volume of terms.

### Definition 3: Optimal Rectangle

A maximal rectangle containing an element (x, y) of a relation R is called **optimal** if it produces a maximal gain with respect to other maximal rectangles containing (x, y).

Figure 2.2(a) presents an example of a relation R, and Figures 2.2(b), 2.2(c) and 2.2(d) represent three maximal rectangles containing the element (y,3). The corresponding gains are 1, 0 e -1. Therefore, the optimal rectangle containing (y,3) of R is the rectangle of Figure 2.2(b).

### Definition 4: Rectangular Graph

Let "≤≤" be a relation defined over a set of rectangles of a binary relation *R*, as follows:

$\forall$ (A1,B1) e (A2, B2) two rectangles of R:

$(A_1, B_1) \leq\leq (A_2, B_2) \Leftrightarrow A_1 \subseteq A_2$ and $B_2 \subseteq B_1$.

We call *(R, ≤≤)* a **Rectangular Graph**.

Note that "≤≤" defines a partial order over the set of rectangles.

### Definition 5: Lattice

An ordered set *(R, <)* is called a **lattice** if each subset $X \subseteq R$ has a minimal upper bound and a maximal lower bound.

### Proposition 1:

Let $R \subseteq E \times G$ be a binary relation. *The set of optimal rectangles of R, ordered by "≤≤" (R, ≤≤) is a lattice with a lower bound ($\emptyset$, G) and an upper bound (E, $\emptyset$).*

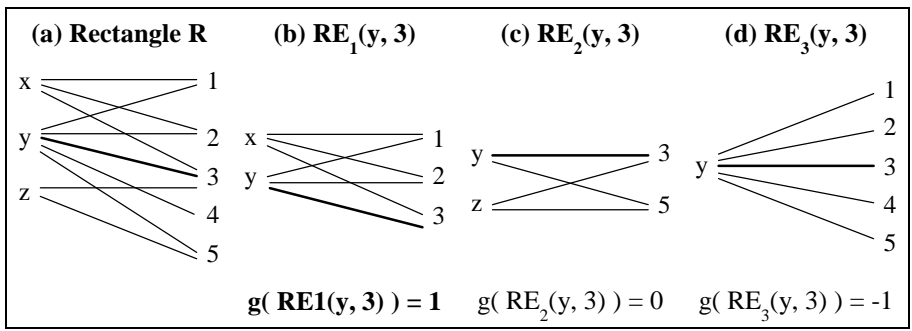

**Figure 2.2: Finding the optimal rectangle**

## 2.1 Semantic Relations

Semantic relations are the basis for the transformation of a rectangular graph in a rectangular thesaurus. They connect the nodes of a rectangular thesaurus in order to find more generic, more specific, or other related rectangles. There are three kinds of semantic relations in a rectangular thesaurus: hierarchical relations (generic and specific), equivalence relations (synonyms and pseudo-synonyms), and neighborhood relations.

The hierarchical relations are based on the following definition:

**Definition 6: Hierarchical relation**

Let $RE_i = (A_i, B_i)$ and $Re = (A_j, B_j)$ be two optimal rectangles of a relation $R$. $RE_i$ is a **generic** of $RE_j$ iff:

$$(A_i, B_i) \leq\leq (A_j, B_j)$$

If the domains are terms and the co-domain are document identifiers, then $A_i$ indexes more documents than $A_j$. The rectangles of interest in this paper will all be of this format, i.e. relating terms to documents. Each term in a rectangle is the representative of an equivalence relation of synonyms. The (representative) terms in the rectangle are called pseudo-synonyms. Note that two terms are pseudo-synonyms if they index the same set of documents.

Figure 2.3 illustrates a hierarchy of three rectangles.

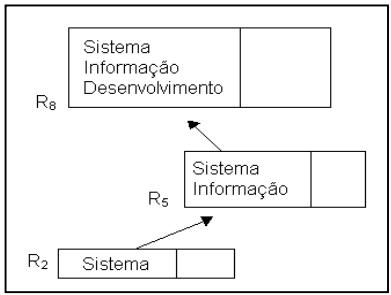

**Figure 2.3: Example of a hierarchy of rectangles**

Figure 2.4 shows the terms *Information, Retrieval* and *Document* as pseudo-synonyms representing: *fact* and *data, search* and *query,* and *report, form* and *article,* respectively.

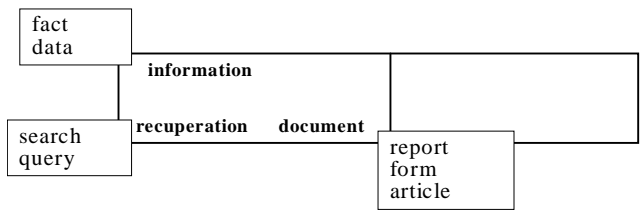

**Figure 2.4: Example of synonyms and pseudo-synonyms**

Non hierarchical relations between rectangles occur when two rectangles have some information in common but none of them is a generic of the other **[12].**

**Definition 8: Non-hierarchical relation**

Let $RE_i = (A_i, B_i)$ and $RE_j = (A_j, B_j)$ be two optimal rectangles of R. $RE_i$ is a **neighbor** $RE_j$ if and only if the following conditions hold:

$$A_i \cap A_j \neq \emptyset \text{ or } B_i \cap B_j \neq \emptyset \quad \text{and}$$

$$(A_i, B_i) \not\leq\leq (A_j, B_j) \text{ and } (A_j, B_j) \not\leq\leq (A_i, B_i)$$

Figure 2.5 shows two neighbor rectangles.

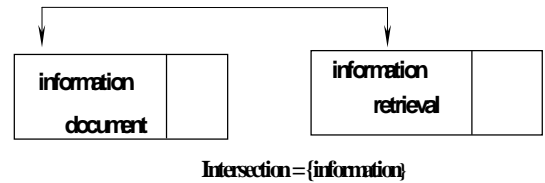

**Figure 2.5: Neighbor rectangles**

Finally we define a **Rectangular Thesaurus** as a graph where the nodes are a set of optimal rectangles an the edges are the semantic relations defined above.

## 2.2 Multilingual Rectangular Thesaurus

We extend the definition of multilingual thesaurus of Sosoaga [16] in order to include the notion of contexts that allow the elimination of ambiguities.

A **multilingual thesaurus** is a classification system defined as

$$MTh = (V, n, r; L_1, C_1, t_1, L_2, C_2, t_2, ..., L_k, C_k, t_k)$$

composed of a unique set of abstract concepts (V), a set of lexicons $\{L_1,.., L_n\}$, a set of contexts $\{C_1, .. , C_k\}$ a set of functions $t_i: L_i \times C_i \rightarrow V$ which associate to each term of the lexicon in a given context a unique abstract term. The hierarchical and non-hierarchical relationships are given by n (narrower term) and r (related term).

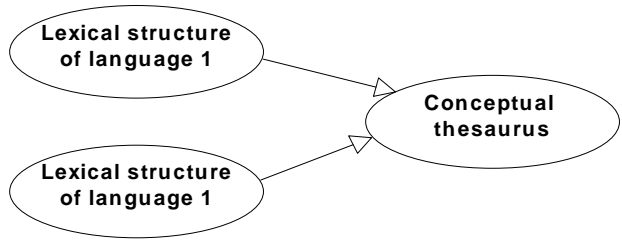

**Figure 2.6: Multilingual Thesaurus**

Therefore, both n and r are subsets of $V \times V$. Figure 2.6 shows a diagram representing a multilingual thesaurus.

A **rectangular multilingual thesaurus** is a multilingual thesaurus where the elements of V are optimal rectangles and the relations n and r follow the definitions of hierarchical and non-hierarchical semantic relations between rectangles, defined above.

In order to construct rectangles with representative terms, we can decompose each function t in two parts $t_0$ and $t_1$, with $t(x, c) = t_1(t_0(x, c))$. The function $t_0$ is responsible for the selection of the canonical representative between the synonyms in a given context and $t_1$ is an injective function that determines the abstract term associated with the original term in a given language.

## 2.3 Simplified representation of a rectangular thesaurus

The hierarchical connected nodes of a rectangular thesaurus contain a lot of redundancy, since if $(A_1,B_1) \leq\leq (A_2, B_2)$ we now that $A_1 \subseteq A_2$ and $B_2 \subseteq B_1$. For large document bases this become prohibitive. For this reason we apply the simplification algorithm of [6] and if $(A_1,B_1) \leq\leq (A_2, B_2)$ the rectangle $(A_2, B_2)$ is represented by $(A_2-A_1, B_1-B_2)$, without loss of information content.

## 3. Construction of a Rectangular Multilingual Thesaurus

The construction of a Rectangular Multilingual Thesaurus is done in three steps:

- Term extraction from one or more documents and determination of the abstract concepts, using a monolingual dictionary (semi-automatic indexing);
- Generation of one or more optimal rectangles;
- Optimal insertion of the new rectangles into the existing thesaurus.

### 3.1 Semi-automatic indexing

The construction of a rectangular thesaurus referencing a set of electronic documents in a natural language begins with the extraction of relevant terms contained in the document. Our semi-automatic method allows the user to eliminate ambiguities interactively.

The first step consists of words selection, stopword elimination and, for significant words, finding of the abstract term. As shown in Figure 3.1, two dictionaries are used for this step. The first, a dictionary of term variations, contains all lexical variations of words in a language and determines its normal form. Compound words, such as 'Data Base' or 'Operating System' must be considered as single terms, since in other language they can be written as single word (e.g. 'Datenbank' and 'Betriebssystem' in German) and should be associated to a single abstract term code. These are identified by a look-ahead step if the dictionary identifies a candidate compound word.

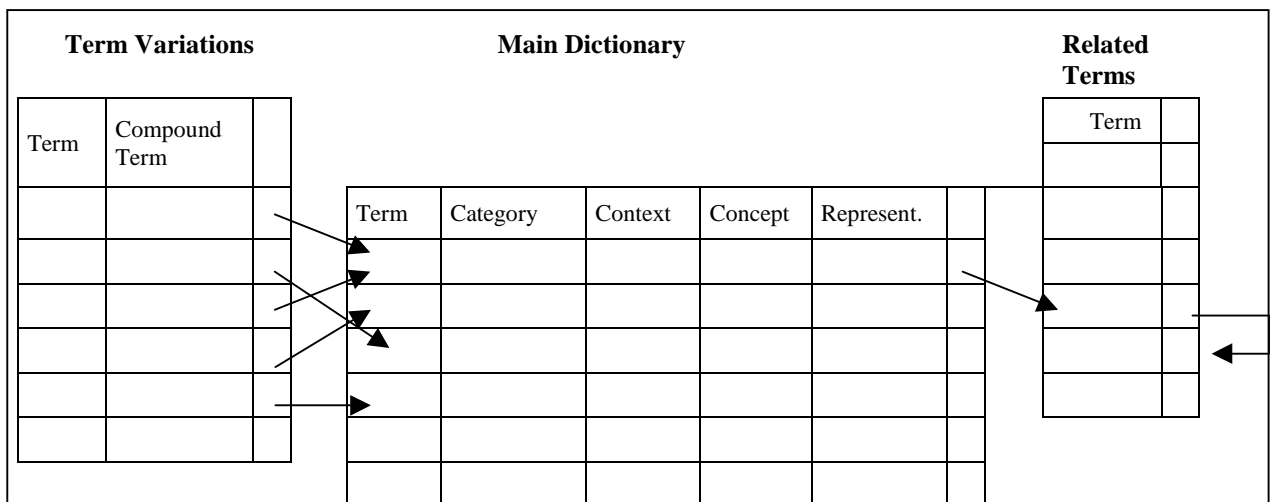

**Figure 3.1: Dictionaries**

Having found the normal form of a term, the main dictionary is then used to find the abstract language-independent term, depending on the context.

In the main dictionary, the column "Representative" and the list of "Related terms" will be used in the construction of the thesaurus in a given language for query formulation (see section 4 below).

In the next subsection we show how, based on Bruandet's method, the association of (abstract) terms is determined. The co-occurrence of pairs of terms is analyzed thus allowing the determination of the binding force between the terms and the choice of candidates for rectangle construction. Since the complete analysis is done at the abstract concept level, term variations and synonyms are considered to represent the same concept.

### 3.2 Measure of the term-term connections

Let **C** be the set of concepts associated with the terms extracted from a set of documents. We want do determine a measure of the contextual proximity of two concepts. This measure is based on the works of Attar e Fraenkel **[2]**. The i-th occurrence of a term[1] *x* of a vocabulary **T**, written as $w_x(i)$, is defined by:

$$w_x(i) = \langle ND_x(i), NF_x(i), NT_x(i) \rangle$$

where $ND_x(i)$ is the number of the document containing x; $NF_x(i)$ the number of the phrase in the document; $NT_x(i)$ the position of the term in the phrase.

For each pair of terms *(x, y)*, we define the distance **d** between the i-th occurrence of *x* and the j-th occurrence of *y*:

$$d(w_x(i), w_y(j)) = \begin{cases} |NT_x(i) - NT_y(j)| & if \begin{cases} ND_x(i) = ND_y(i) \\ and \\ NF_x(i) = NF_y(i) \end{cases} \\ 0, otherwise \end{cases}$$

The binding force between two terms is a function **F** given by the inverse of the distance between the terms:

$$F(w_x(i), w_y(j)) = \begin{cases} \dfrac{1}{d(w_x(i), w_y(j))} & if \ 0 < d(w_x(i), w_y(j)) \le t \\ 0, otherwise \end{cases}$$

**t** *is an experimentally obtained constant defined below*

It is important to consider the grammatical category of the terms. In most cases the categories of interest are nouns, adjectives and verbs. For each language we can determine the distance between the grammatical categories, and this distance will be used to determine the constant **t** in the formula above.

$$t(x, y) = DIST[CAT(x), CAT(y)]$$

*where* DIST[x, y] *is the distance between the grammatical categories of x and y.*

The grammatical categories considered and the values of the function DIST[] can be modified should the application exhibit special needs.

Let b(x, y) be the sum of the binding forces between two terms x and y of a document

$$b(x, y) = \sum_i \sum_j F(w_x(i), w_y(j))$$

and f(x, y) be the number of occurrences of the pair (x, y) in the document. We define the **association measure** between two terms as

$$M(x, y) = \mathbf{k(x, y)} \times \frac{b(x, y)}{f(x, y)} \quad where$$

$$\mathbf{k(x, y)} = \frac{(f(x, y) - 1)^n}{f(x, y)^n}$$

The correction factor **k** can be used if we want to avoid that pairs of terms that occur only once be included in the association. The parameter **n** is defined by experimentation and a typical value is 2 [5]. Higher values of **n** give more weight for pairs occurring frequently and discard pairs with few occurrences.

---

[1] We use the word 'term' as a synonym of 'concept' in the following discussion.

We show in Figure 3.1 an example of the computing of M(t1,t2) using a text on object orientation.

| Term1(t1) | Term2(t2) | b(t1,t2) | f(t1,t2) | b(t1,t2)/ f(t1,t2) | k | M (t1,t2) |
|---|---|---|---|---|---|---|
| **Orientado** | **Objeto** | **4,25** | **9** | **0,47** | **0,70** | **0,33** |
| **Análise** | **Orientado** | **3** | **3** | **1** | **0,29** | **0,29** |
| **Desenvolvimento** | **Software** | **2** | **4** | **0,5** | **0,42** | **0,21** |
| **Orientação** | **Objeto** | **2** | **4** | **0,5** | **0,42** | **0,21** |
| **Biblioteca** | **Classe** | **1,5** | **3** | **0,5** | **0,29** | **0,14** |
| **Mudança** | **Cultural** | **2** | **2** | **1** | **0,12** | **0,12** |
| **metodologia** | **Análise** | **1,16** | **3** | **0,39** | **0,29** | **0,11** |
| **conceito** | **Objeto** | **1,11** | **3** | **0,37** | **0,29** | **0,11** |
| **ambiente** | **Orientado** | **1,11** | **3** | **0,37** | **0,29** | **0,11** |
| análise | Objeto | 1 | 3 | 0,33 | 0,29 | 0,09 |
| banco | Dados | 1 | 2 | 0,5 | 0,12 | 0,06 |
| desenvolvimento | Objeto | 0,45 | 2 | 0,23 | 0,12 | 0,02 |

**Table 3.1: Examples of computation of association measures**

### 3.3 Generation of optimal rectangles

The values of the association measure M are stored in a binary term-term matrix. Values of M less than a specific constant are set to 0 while the others are set to 1. Note that the matrix is symmetrical and the diagonal can be disregarded. This fact can be used to better choose how to store the information.

After selecting the relevant terms for each document, based on the association measure, the original binary relation between terms and documents is reduced to its significant terms only.

We apply the method of rectangular decomposition of a binary relation R [6] in order to obtain the optimal rectangles. The main steps in this method are the following:
1. Decompose R in n elementary relations $ER_1, .., ER_n$
2. Select, in parallel, for each elementary relation $ER_i$ the set of optimal rectangles;
3. Select, for each $ER_i$, the minimum optimal rectangles that cover the relation;
4. Eliminate as many redundant elements as possible in the optimal rectangles.

Although the general maximal rectangle selection problem is NP-hard, the heuristics given in [6] are $O(n^3)$.

### 3.4 Updating the rectangular thesaurus

Each rectangle obtained in the previous steps relates a set of terms to a set of documents. If we are processing a single document, one rectangle is generated with the significant terms indexing that document. We must now insert the new rectangles in the existing abstract rectangular thesaurus.

Figure 3.2 shows the rectangular thesaurus for a document base, where the abstract terms of the domains of the rectangles have been exchanged by natural ones. Since it is in simplified form, term redundancy has been eliminated.

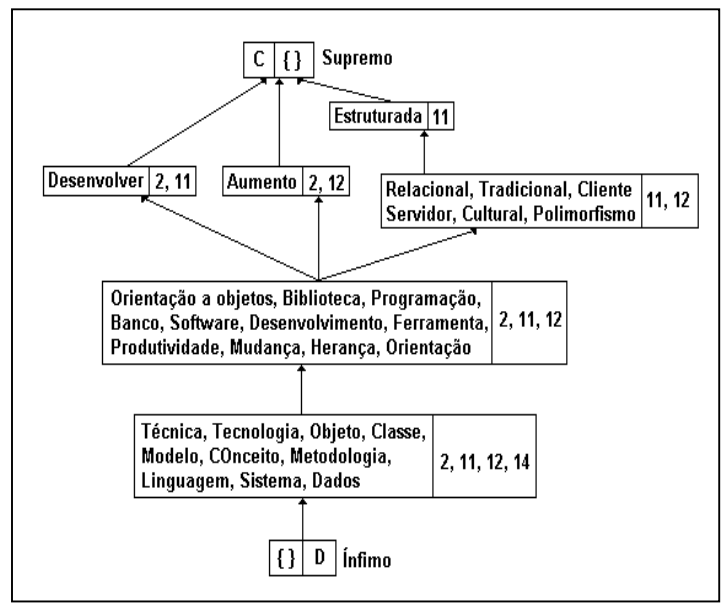

**Figure 3.2: Document thesaurus**

Note that in a rectangular thesaurus, we can identify several cardinality levels, due to the cardinality of the rectangle domain. These levels vary from 0 to n, where n is the total number of terms in the thesaurus. Each new rectangle must be placed in the corresponding level. In the example, the second level has cardinality 10 and the third level has cardinality 21, since 11 terms have been added to the level below.

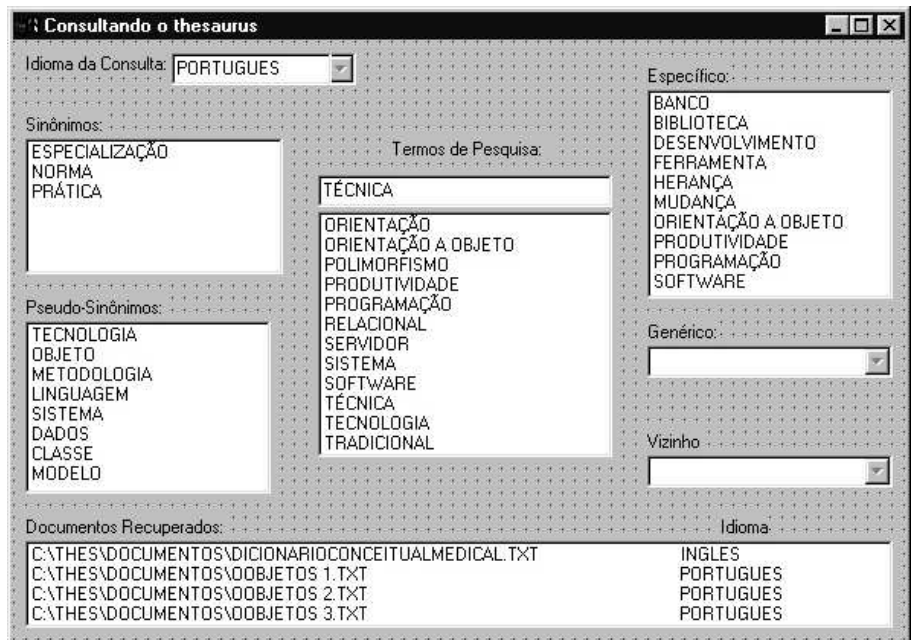

The following algorithm, from **[14],** provides the insertion of a new rectangle in an existing rectangular thesaurus. We consider the thesaurus in its original definition, without simplification. Considering the simplification is straightforward.

*1. Check if the cardinality level of the new rectangle exists in the thesaurus*
   *1.1. If it does not exist, create the new level for the rectangle*
   *1.2. else*
      *1.2.1. If the domain of the new rectangle coincides with an existing rectangle, then add the new document to the co-domain of the existing rectangle*
      *else insert the new rectangle in the level*
*2. If a new rectangle has been created, establish the hierarchical connections, searching for the higher level rectangles containing the new terms.*
*3. New terms not occurring in the supremum, are inserted there*
*4. If the descendants of the new rectangle are empty, connect it to the infimum*

## 4. Information Retrieval

The purpose of an information retrieval system is to return to the user a set of documents that match the keywords expressed in the query. In our system we present an interface, using the user's preferred language, representing the thesaurus of concepts occurring in the document database. This thesaurus includes synonyms, related terms and hierarchical relations.

The dictionary-like form of the interface is not as user-friendly as generic natural language interfaces, but it avoids the well known problems with natural languages and guarantees fast access to the documents searched for. As reported by LYCOS, typical user queries are only two or three words long **[17].**

Figure 4.1 shows the protoype's interface **[15]** with terms defined in Portuguese.

In a rectangular thesaurus, the retrieval process consists of finding a rectangle Ri = Ci x Di, such that Ci is a minimal domain containing the set of terms from the query Q. If Ci ≠ Q the user can receive feedback from the system concerning other terms which index the retrieved documents. This fact is identified as a Galois connection in [6]. Note that we can obtain several rectangles matching the query. On the other hand, the user can eliminate some terms from the query in order to obtain more documents.

As can be seen in the figure, the prototype allows one to chose a language and as he/she is selecting the terms, the systems lists the corresponding documents.

## 5. Conclusion

Most work on thesaurus construction uses automatic indexing of a given set of documents **[5] [6] [18] [19]** and, in the case of a multilingual framework, use machine translation techniques applied on the query **[17].** In **[14]** an incremental version of the approach on automatic generation of rectangular thesauri of Bruandet and Gammoudi has been developed. The main contribution of this paper is to integrate the incremental approach with a dictionary-based multilingual indexing and information retrieval and to propose an interactive ambiguity resolution. This approach eliminates problems of automatic indexing, linguistic variations of a single concept and restrictions of monolingual systems. Furthermore the problem of terms composed of more than one word has been solved with a look-ahead algorithm for candidate words found in the dictionary.

It is clear that the interaction with the user is very time-consuming. But, it seems to be a good trade-off between the precision of manual thesaurus construction and the efficiency of automatic systems. With an 'apply to all' option on can avoid repeating conflict resolutions.

The indexing approach used has two drawbacks which have not been solved up to now:

1. The hierarchical relations in the rectangular thesaurus are purely syntactic, i.e. more composed terms are more specific that less composed ones;
2. The relevance analysis of pairs of term completely eliminates single terms of phrases. As a result, a term can occur very frequently in a document but, if it appears associated to distinct terms in different phrases, it could be eliminated as an indexing candidate.

One possible solution to the first problem is to use of a richer dictionary for term analysis. If the dictionary has thesaurus-like information, this can be used for the construction of the rectangular term-document thesaurus. For the second problem, a word counter, running in parallel with the indexing, could help the inclusion of single terms.